\documentclass[preprint2]{aastex}

\shorttitle{Bright Milagro sources in the Cygnus Region}
\shortauthors{Bonamente et al.}

\usepackage{lineno}

\begin{document}

%%%%%\linenumbers


\title{Spectrum and Morphology of the Two Brightest Milagro Sources in the
       Cygnus Region: MGRO J2019+37 and MGRO J2031+41.}


\author{
  A.~A.~Abdo,\altaffilmark{\ref{msu},\ref{nrl}}
  B.~T.~Allen,\altaffilmark{\ref{uci},\ref{cfa}}
  T.~Aune,\altaffilmark{\ref{ucsc}}
  D.~Berley,\altaffilmark{\ref{umcp}}
  E.~Bonamente,\altaffilmark{\ref{mtu}}
  G.~E.~Christopher,\altaffilmark{\ref{nyu},\ref{ngc}}
  T.~DeYoung,\altaffilmark{\ref{psu}}
  B.~L.~Dingus,\altaffilmark{\ref{lanl}}
  R.~W.~Ellsworth,\altaffilmark{\ref{georgemason}}
  J.~G.~Galbraith-Frew,\altaffilmark{\ref{mtu}}
  M.~M.~Gonzalez,\altaffilmark{\ref{ida}}
  J.~A.~Goodman,\altaffilmark{\ref{umcp}}
  C.~M.~Hoffman,\altaffilmark{\ref{lanl}}
  P.~H.~H\"untemeyer,\altaffilmark{\ref{mtu}}
  B.~E.~Kolterman,\altaffilmark{\ref{nyu}}
  J.~T.~Linnemann,\altaffilmark{\ref{msu}}
  J.~E.~McEnery,\altaffilmark{\ref{gsfc}}
  A.~I.~Mincer,\altaffilmark{\ref{nyu}}
  T.~Morgan,\altaffilmark{\ref{unh}}
  P.~Nemethy,\altaffilmark{\ref{nyu}}
  J.~Pretz,\altaffilmark{\ref{lanl}}
  J.~M.~Ryan,\altaffilmark{\ref{unh}}
  P.~M.~Saz~Parkinson,\altaffilmark{\ref{ucsc}}
  G.~Sinnis,\altaffilmark{\ref{lanl}}
  A.~J.~Smith,\altaffilmark{\ref{umcp}}
  V.~Vasileiou,\altaffilmark{\ref{umcp},\ref{cresst}}
  G.~P.~Walker,\altaffilmark{\ref{lanl},\ref{nst}}
  D.~A.~Williams\altaffilmark{\ref{ucsc}}
  and
  G.~B.~Yodh\altaffilmark{\ref{uci}}}

\altaffiltext{1}{\label{msu} Department of Physics and Astronomy, Michigan State University, 3245 BioMedical Physical Sciences Building, East Lansing, MI 48824}
\altaffiltext{2}{\label{nrl} Current address: Space Science Division, Naval Research Laboratory, Washington, DC 20375}
\altaffiltext{3}{\label{uci} Department of Physics and Astronomy, University of California, Irvine, CA 92697}
\altaffiltext{4}{\label{cfa} Current address: Harvard-Smithsonian Center for Astrophysics, Cambridge, MA 02138}
\altaffiltext{5}{\label{ucsc} Santa Cruz Institute for Particle Physics, University of California, 1156 High Street, Santa Cruz, CA 95064}
\altaffiltext{6}{\label{umcp} Department of Physics, University of Maryland, College Park, MD 20742}
\altaffiltext{7}{\label{mtu} Department of Physics, Michigan Technological University, Houghton, MI 49931}
\altaffiltext{8}{\label{nyu} Department of Physics, New York University, 4 Washington Place, New York, NY 10003}
\altaffiltext{9}{\label{ngc} Current address: Department of Physics, Brown University, Providence, RI 02912, USA}
\altaffiltext{10}{\label{psu} Department of Physics, Pennsylvania State University, University Park, PA 16802}
\altaffiltext{11}{\label{lanl} Group P-23, Los Alamos National Laboratory, P.O. Box 1663, Los Alamos, NM 87545}
\altaffiltext{12}{\label{georgemason} Department of Physics and Astronomy, George Mason University, 4400 University Drive, Fairfax, VA 22030}
\altaffiltext{13}{\label{ida} Instituto de Astronom\'ia, Universidad Nacional Aut\'onoma de M\'exico,
  74D.F., M\'exico, 04510}
\altaffiltext{14}{\label{gsfc} NASA Goddard Space Flight Center, Greenbelt, MD 20771}
\altaffiltext{15}{\label{unh} Department of Physics, University of New Hampshire, Morse Hall, Durham, NH 03824} %-3525}
\altaffiltext{16}{\label{cresst} CRESST NASA/Goddard Space Flight Center, MD 20771 and University of Maryland, Baltimore County, MD 21250}
\altaffiltext{17}{\label{nst} Current address: National Security Technologies, Las Vegas, NV 89102}

\email{ebonamen@mtu.edu, petra@mtu.edu}

\begin{abstract}
\noindent
The Cygnus region is a very bright and complex portion of the TeV sky, host to unidentified sources
and a diffuse excess with respect to conventional cosmic-ray propagation models.
Two of the brightest TeV sources, MGRO J2019+37 and MGRO J2031+41, are
analyzed using Milagro data with a new technique, and their emission is tested under
two different spectral assumptions: a power law and a power law with an exponential cutoff.
The new analysis technique is based on an energy estimator that uses the fraction
of photomultiplier tubes in the observatory that detect the extensive air shower.
The photon spectrum is measured in the range 1 to 200 TeV using the last 3 years of Milagro data
(2005-2008), with the detector in its final configuration.
MGRO J2019+37 is detected with a significance of 12.3 standard deviations ($\sigma$), and is better fit
by a power law with an exponential cutoff than by a simple power law, with a probability $>98$\% (F-test).
The best-fitting parameters for the power law with exponential cutoff model are a normalization at 10 TeV
of $7^{+5}_{-2}\times10^{-10}$ $\mathrm{s^{-1}\: m^{-2}\: TeV^{-1}}$, a spectral index of $2.0^{+0.5}_{-1.0}$
and a cutoff energy of $29^{+50}_{-16}$ TeV.
MGRO J2031+41 is detected with a significance of 7.3$\sigma$, with no evidence of a cutoff.
The best-fitting parameters for a power law are a normalization of $2.4^{+0.6}_{-0.5}\times10^{-10}$
$\mathrm{s^{-1}\: m^{-2}\: TeV^{-1}}$ and a spectral index of $3.08^{+0.19}_{-0.17}$.
The overall flux is subject to an $\sim$30\% systematic uncertainty.
The systematic uncertainty on the power law indices is $\sim$0.1.
A comparison with previous results from TeV J2032+4130, MGRO J2031+41 and MGRO J2019+37 is also presented.

\end{abstract}


% INTRODUCTION
\section{Introduction}

The Cygnus region is a part of our Galaxy of active massive star formation and destruction, and has been 
studied over a broad range of wavelengths, including radio, X-ray, GeV and TeV gamma-ray, as well as in cosmic rays.
From GeV up to multi-TeV energies the Cygnus region is the brightest diffuse gamma-ray source in the northern 
hemisphere \citep{egret1997}.

One of the challenges in analyzing the Cygnus region at TeV energies, is the proper separation of the gamma-ray
flux that is attributed to the point or extended sources in the region or to the diffuse emission. 
Previous Milagro analyses computed the diffuse emission from the region using two alternative methods to
isolate the contribution from the resolved sources \citep{milagro2007a,milagro2008}, and found that at TeV
energies the flux is still in excess with respect to the predicted flux from the GALPROP model \citep{galprop}.
Milagro also published the discovery of two unidentified TeV sources in the region, MGRO J2019+37 and MGRO J2031+41
\citep{milagro2007b,milagro2009}.
The location of MGRO J2019+37 was found to be consistent with two EGRET sources, while the best fit position for
MGRO J2031+41 was near two EGRET sources and the HEGRA unidentified source TeV J2032+4130.
In a correlation study connecting the TeV sources discovered by Milagro with sources detected above 10$\sigma$
(the so-called Bright Source List, BSL) by the Fermi Large Area Telescope (LAT), the two aforementioned brightest
Milagro sources in the region were associated with GeV pulsars \citep{fermi2009c,milagro2009}.
MGRO J2019+37 and MGRO J2031+41 are currently associated with two and one pulsars identified by the Fermi LAT,
respectively \citep{fermi2011c}.
Milagro also detected two candidate sources in the Cygnus region ($l\in[65^\circ, 85^\circ]$ and
$b\in[-2^\circ,+2^\circ]$) \citep{milagro2009}.

Recently, several collaborations have presented new surveys of the Cygnus region,
confirming the complexity of the region and showing the highly structured diffuse emission.
The TeV emission from the position of MGRO J2019+37 has been confirmed by the VERITAS experiment
\citep{veritas2011}.
The VERITAS spatial counts map shows a clear structure associated with MGRO J2019+37, a region of extended
emission which seems to be produced by previously unresolved sources.
At lower energies, the Fermi LAT collaboration has also recently published new results on the diffuse emission
from the Cygnus region.
An extended source, the so-called Cocoon, overlapping with MGRO J2031+41, has been found
and its emission has been explained by freshly accelerated cosmic rays, trapped in a shell of photon-dominated
emission formed by stellar winds and supernovae, as shown by mid-infrared observations
\citep{fermi2011b}.
The spectrum from this region is hard, extending up to 100 GeV with no evidence of softening, and could be
explained as the result of pulsar-accelerated particles or as an active super-bubble.
The average diffuse emission from the region has also been analyzed at MeV to GeV energies \citep{fermi2011a}.
Despite the very rich source population, this emission is similar to that of the local interstellar space, and there
is no evidence that it is necessary to include an extra contribution in the model, resembling the diffuse
excess previously measured by Milagro \citep{milagro2007a,milagro2008}.
Most recently, the ARGO-YBJ collaboration presented the results of a survey of the Cygnus region in the energy
range of 600 GeV to 7 TeV.
MGRO J2031+41 is observed with a significance of 6.4$\sigma$ and a flux consistent with previous
Milagro results, but there is no significant detection of MGRO J2019+37 \citep{argo2012}.

Here, we present a new analysis of the last three years of data collected with the Milagro experiment (2005-2008).
An improved gamma-hadron separation and a newly-developed technique are exploited to reconstruct the energy
spectra of gamma rays from the sources in the Cygnus region.
The best fits for the spectra of the two brightest Milagro sources are presented:
MGRO J2019+37, a source with a post-trials significance in excess of 12$\sigma$
between 1 and 100 TeV, and MGRO J2031+41 a source with a post-trials significance in excess of
7$\sigma$ \citep{milagro2007b}.
We compare our spectra with results from the HEGRA, MAGIC, Whipple and ARGO-YBJ experiments
\citep{hegra2005,magic2008,whipple2004,argo2012}.

% ANALYSIS
\section{Analysis Technique}\label{s.analysis}

The Milagro detector was located in the Jemez Mountains in New Mexico, at an altitude of 2,630 m a.l.s.
It was operated from 2001 to 2008, and, in its final configuration, consisted of two components:
(1) a central pond ($60\times80$ $\mathrm{m^2}$, 8 m deep) with two photomultiplier tube (PMT) layers,
a shallow {\it air shower} layer consisting of 450 PMTs, and a deep {\it muon layer} consisting of 273 PMTs and
(2) an array of 175 single-PMT tank detectors surrounding the pond covering an area of 40,000 $\mathrm{m^2}$
\citep{milagro2001,milagro2003}.

A detailed description of the analysis method and the parameters used here can be found in the
Milagro paper on the spectral measurement of the Crab Nebula \citep{milagro2011}.
Systematic uncertainties of this method are estimated to be $\sim$30\% on the overall flux and $\sim$0.1 on
the spectral index.

The new element in this analysis, with respect to the previous approach \citep{milagro2007b,milagro2009},
is the introduction of an estimator, $\mathcal{F}$, for the energy of the primary particle initiating the extensive
air shower, based on the number of PMTs hit for each event.
The $\mathcal{F}$ parameter is defined as the sum of two fractions:
\begin{equation}
\mathcal{F}=\frac{N_{A.S.}}{N_{A.S.}^{live}}+\frac{N_{T.A.}}{N_{T.A.}^{live}},
\label{F}
\end{equation}
where $N_{A.S.}$ and $N_{T.A.}$ are the number of PMTs detecting the event, while $N^{live}_{A.S.}$ and
$N^{live}_{T.A.}$ are the number of functional PMTs in the air shower layer and in the tank array, respectively.
Because the typical energy resolution of Milagro is quite broad (50-100\%), and the median energy associated with
a given $\mathcal{F}$ value is dependent on the spectral assumption for the observed source, the fit is performed
in the measured $\mathcal{F}$ space rather than in the energy space.

The reconstructed Milagro data contain information about the direction and $\mathcal{F}$ value of air shower events.
According to their $\mathcal{F}$ value, events are sorted into 9 bins ($0.2\le \mathcal{F} \le 2$),
resulting in a total of 9 signal skymaps, binned in $0.1^{\circ}\times0.1^{\circ}$ pixels.
For each $\mathcal{F}$ bin, background maps are calculated using the Direct Integration method \citep{milagro2011}.
The two brightest regions of interest in the sky, a $2^{\circ}\times2^{\circ}$ area around the position of
the Crab Nebula and a band around the Galactic plane ($-2.5^{\circ}<b<2.5^{\circ}$) are excluded when calculating
the background \citep{milagro2011}.
Rather than discriminating between gamma-ray and cosmic-ray initiated air showers, a weight is
applied to all measured events, in both signal and background maps, where gamma-ray like events are assigned a
higher weight than cosmic-ray like events \citep{milagro2011}.
The gamma-ray excess with respect to the estimated background is calculated as the difference between
signal weights and background weights, and is filled into the so-called excess map.
Based on these excess maps we compute the $\mathcal{F}$ distributions used in the energy fit.

Excess maps are smoothed according to the detector angular resolution (or point spread function, PSF),
which is a function of $\mathcal{F}$ and varies between $0.3^{\circ}$ and $0.7^{\circ}$.
The measured $\mathcal{F}$ distribution for the target source in a $0.1^{\circ}\times0.1^{\circ}$ bin, as a result,
is the average excess coming from a PSF-wide region around the nominal source position.
Since the background is measured with the direct integration method removing the Galactic plane ($b>2.5^{\circ}$
and  $b<-2.5^{\circ}$), the measured excess from sources in the Galactic plane includes any other diffuse
or extended emission possibly present in the vicinity of the source.
The Galactic diffuse background was estimated to contribute up to 15\% of the flux at 35 TeV for the weakest
Galactic BSL source \citep{milagro2009}.

We test the emission from the two sources for two different spectral hypotheses:
a power law, defined as
\begin{equation}
\frac{dN}{dE}=N_{10\mathrm{TeV}}\left( \frac{E}{10\;\mathrm{TeV}}\right)^{-\alpha},
\label{e.PL}
\end{equation}
where $N_{10\mathrm{TeV}}$ is the normalization scale set at 10 TeV and $\alpha$ the spectral index,
and a power law with an exponential cutoff, defined as
\begin{equation}
\frac{dN}{dE}=N_{10\mathrm{TeV}}\left( \frac{E}{10\;\mathrm{TeV}}\right)^{-\alpha}\mathrm{exp}\left( -\frac{E}{E_c}\right),
\label{e.CO}
\end{equation}
where $E_c$ is the cutoff energy.

The fit to the data is performed comparing the measured excess to simulations.
First, a set of simulated data is generated varying sensitive parameters (i.e. $N_{10\mathrm{TeV}}$, $\alpha$ and $E_c$).
The best spectral parameters and the corresponding fit probability are then found using a $\chi^2$ minimization,
comparing the measured and the simulated $\mathcal{F}$ distributions.
Uncertainties on fit parameters are computed using 1-$\sigma$ contours of $\chi^{2}$ histograms (as discussed in
sections \ref{s.2019} and \ref{s.2031}).
The uncertainty is defined as the distance between the best fit value and the lower
and upper edges of the 1-$\sigma$ contour.

% RESULTS
\section{Results and Discussion}\label{s.results}
The results presented here are obtained from the analysis of the Milagro data taken with
the detector in its final configuration.
The start date is 2005-10-22 20:39:16 GMT and the stop date is 2008-04-15 00:02:53 GMT, corresponding to a
total of 906 days of observation, resulting in 832 integrated days after data quality cuts.

The positions of MGRO J2019+37 and MGRO J2031+41 are obtained fitting Milagro excess maps with
a 2-dimensional Gaussian function with equal widths in right ascension and declination ($\sigma_{RA}=\sigma_{Dec}$),
plus a constant base to take into account contribution from diffuse emission.
The fit is performed in a square region of $\pm 3^{\circ}$ around the excess.
Table \ref{t.positions} shows the positions used in this paper and a comparison to other surveys.

%MGRO J2019+37
\subsection{MGRO J2019+37 spectrum}\label{s.2019}
\begin{figure}
\begin{center}
\includegraphics[angle=0,width=8cm]{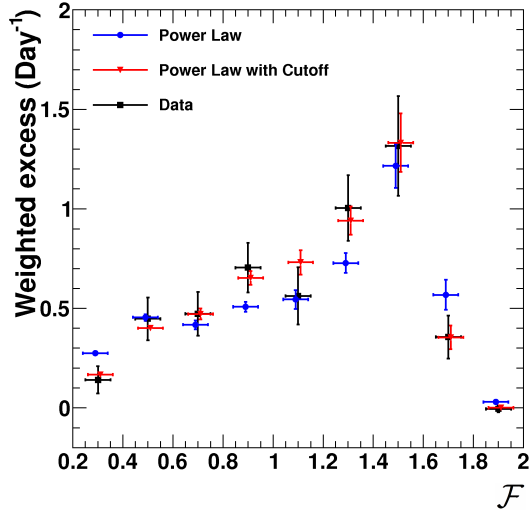}
\end{center}
\caption{MGRO J2019+37: distributions of the parameter used to estimate the photon energy, $\mathcal{F}$.
The spectrum fit is performed in the $\mathcal{F}$ space.
Black squares represent the data. Red triangles and blue circles represent the simulated distributions 
for the best fit assuming a power law with a cutoff and a simple power law, respectively.
The unit on the y axis is the weighted excess per day, the unit on the x axis is the $\mathcal{F}$ value.
\label{f.2019frasor}
}
\end{figure}

In Fig. \ref{f.2019frasor} the $\mathcal{F}$ distribution of data is shown with the two simulated
$\mathcal{F}$ distributions for the best fitting power law and power law with exponential cutoff models.
These distributions show the weighted excess coming from the source position after background
subtraction.
The excess from MGRO J2019+37 is compared to the excess measured from the entire Cygnus region
($65^{\circ}<l< 85{^\circ}$ and $-2^{\circ}<b<2^{\circ}$).
We find that the emission from MGRO J2019+37 accounts for $\sim$12$\%$ of the excess
from the entire region.

\begin{figure}
\begin{center}
\includegraphics[angle=0,width=5cm]{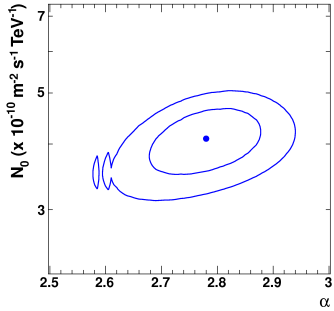}
\end{center}
\caption{MGRO J2019+37: 1 and 2-$\sigma$ C.L. contours for the power law fit. The point indicates the best fit
result (minimum $\chi^2$).
The $\chi^2$ increments for the 1 and 2-$\sigma$ contours are 2.30 and 6.18 respectively.
\label{f.2019contPL}}
\end{figure}

\begin{figure}
\begin{center}
\includegraphics[angle=0,width=7.5cm]{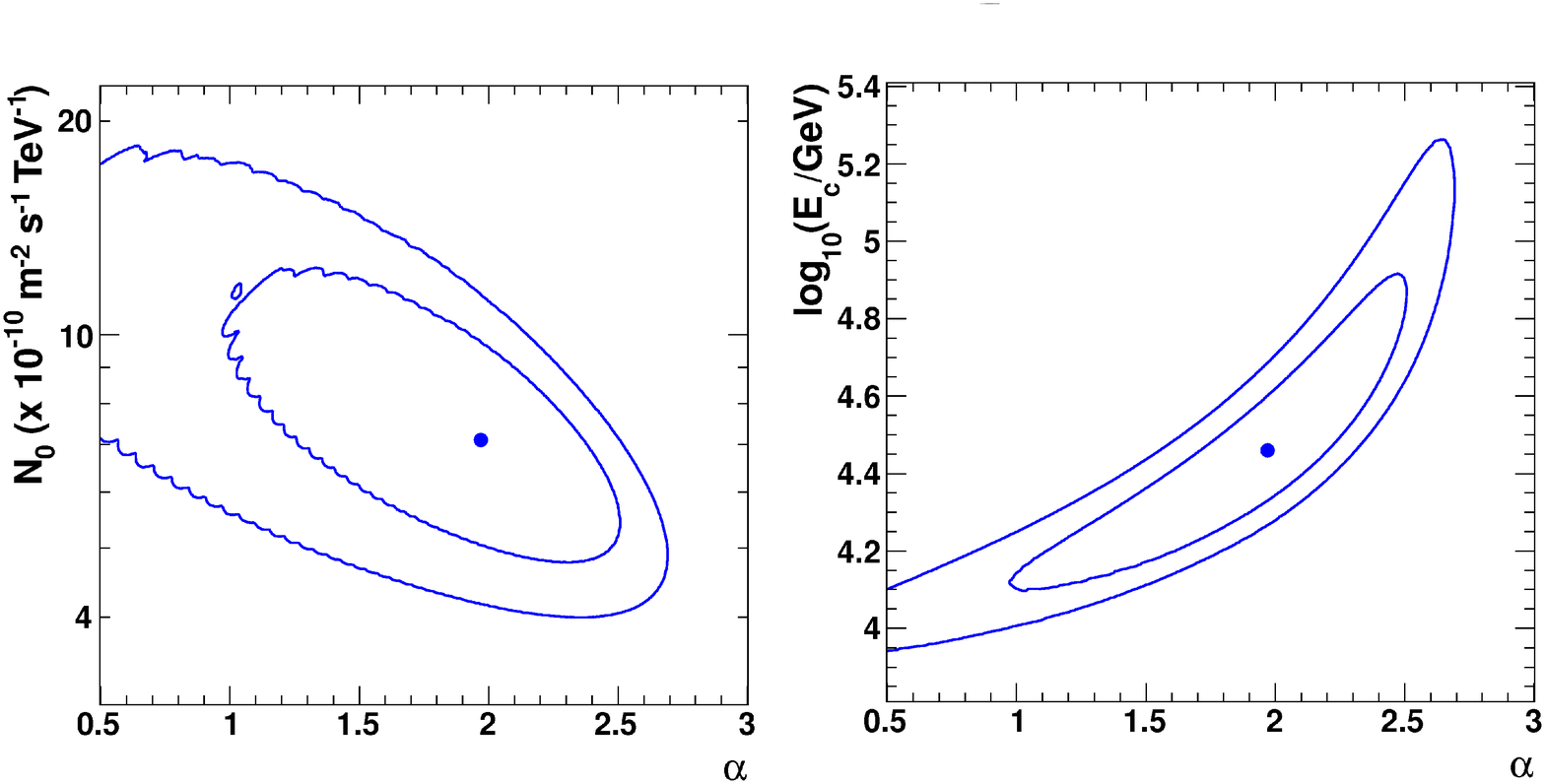}
\end{center}
\caption{MGRO J2019+37: 1 and 2-$\sigma$ C.L. contours for the power law with cutoff fit.
\textit{left panel} - $N_{10\mathrm{TeV}}$ vs $\alpha$ projection. \textit{right panel} - $E_c$ vs $\alpha$ projection.
The points indicate the best fit result (minimum $\chi^2$).
The $\chi^2$ increments for the 1 and 2-$\sigma$ contours are 3.53 and 8.02 respectively.
\label{f.2019contCO}}
\end{figure}

Figs. \ref{f.2019contPL} and \ref{f.2019contCO} show the 1 and 2-$\sigma$ contours from the $\chi^2$
fit for the two tested hypotheses.
1-$\sigma$ contours are used to compute uncertainty on the parameters as described in section \ref{s.analysis}.

\begin{figure}
\begin{center}
\includegraphics[angle=0,width=8cm]{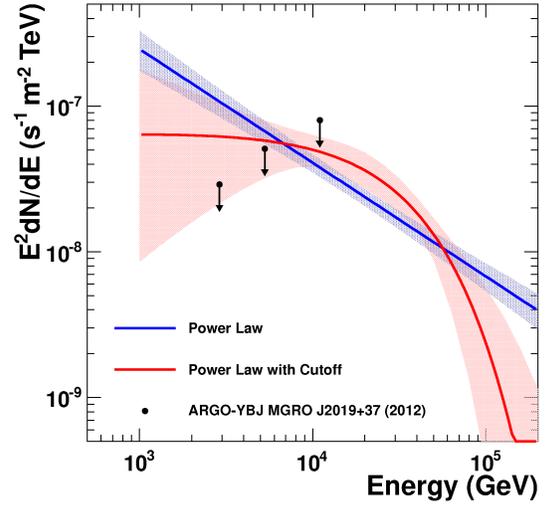}
\end{center}
\caption{MGRO J2019+37: energy spectra.
The best fit is obtained for a power law with cutoff model (in red).
The power law model is also shown (in blue).
The shadowed area represents the 1-$\sigma$ band, obtained by varying the parameters
within the 1-$\sigma$ contour.
ARGO-YBJ 90\% C.L. upper limits for MGRO J2019+37 are shown in black \citep{argo2012}.
\label{f.2019spectra}
}
\end{figure}

The best fit spectra for the two tested hypotheses (Eqs. \ref{e.PL} and \ref{e.CO}) are shown in
Fig. \ref{f.2019spectra}.
The best-fit spectral parameters are summarized in Table \ref{t.results}.

With the power law hypothesis, the best-fit parameters for the normalization and the spectral index
are $N_{10\mathrm{TeV}} = (4.1\pm0.6) \cdot 10^{-10}$ $\mathrm{s^{-1}\: m^{-2}\: TeV^{-1}}$ and $\alpha=2.78\pm0.10$
respectively.
The $\chi^{2}$ is 16.12 for 7 degrees of freedom (d.o.f.), which gives a $\chi^{2}$ fit probability of 0.024.

In the case of the power law with cutoff hypothesis, the best fit parameters are 
$N_{10\mathrm{TeV}}=7^{+5}_{-2}\cdot10^{-10}$ $\mathrm{s^{-1}\: m^{-2}\: TeV^{-1}}$, $\alpha=2.0^{+0.5}_{-1.0}$ and
$E_c=29^{+50}_{-16}$ TeV.
The $\chi^{2}$ is 1.924 (6 d.o.f.), which gives a $\chi^{2}$ fit probability of 0.93.

The $\chi^{2}$ fit probabilities suggest that the power law with a cutoff model ($\chi^{2}$ probability = 93\%)
fits the data better than a simple power law ($\chi^{2}$ probability = 2.4\%).
We use the F-test \citep{ftest} to compute if the improvement in the fit is significant.
For MGRO J2019+37, the simple power law is disfavoured at a C.L. $>98$\%.

Previous Milagro analyses quoted the flux of MGRO J2019+37 at 20 and 35 TeV respectively
\citep{milagro2007b,milagro2009}.
Those values, using a different analysis technique, are in agreement with the results presented here.
An independent analysis \citep{allen2007}, which used a different energy estimator for the particle initiating
the air shower and a different parameter to distinguish between gamma and cosmic rays, also is consistent
with our results.
ARGO-YBJ \citep{argo2012} has not observed any significant emission
from MGRO J2019+37 in the range 600 GeV to 7 TeV.
The explanations put forward in \citep{argo2012} are an insufficient exposure above 5 TeV or a possible time
variability of MGRO J2019+37.
The 90\% C.L. upper limit from ARGO-YBJ is consistent with our best fitting model, a power law with an
exponential cutoff (see Fig. \ref{f.2019spectra}).
On the other hand, the simple power law model ($\alpha=2.78\pm0.10$), already disfavored by the F-test,
does not agree with the ARGO-YBJ results.

%MGRO J2031+41
\subsection{MGRO J2031+41 spectrum}\label{s.2031}
\begin{figure}
\begin{center}
\includegraphics[angle=0,width=8cm]{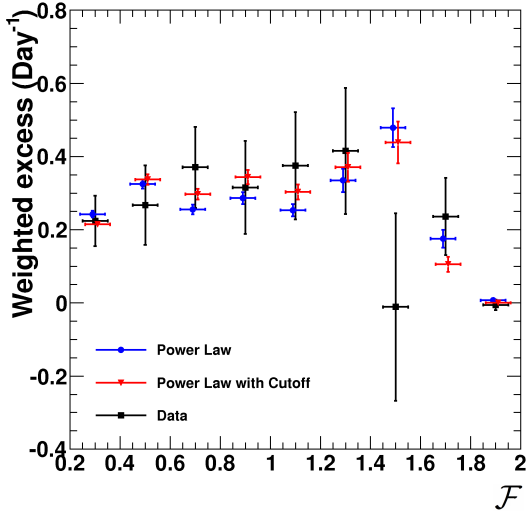}
\end{center}
\caption{MGRO J2031+41: distributions of the parameter used to estimate the photon energy, $\mathcal{F}$.
The spectrum fit is performed in the $\mathcal{F}$ space.
Black squares represent the data. Red triangles and blue circles represent the simulated distributions 
for the best fit assuming a power law with a cutoff and a simple power law, respectively.
The unit on the y axis is the weighted excess per day, the unit on the x axis is the $\mathcal{F}$ value.
\label{f.2031frasor}
}
\end{figure}

The $\mathcal{F}$ distribution from data and those simulated for the two best fit models are shown in
Fig. \ref{f.2031frasor}.
MGRO J2031+41 accounts for $\sim$6$\%$ of the excess from the entire Cygnus region.

\begin{figure}
\begin{center}
\includegraphics[angle=0,width=5cm]{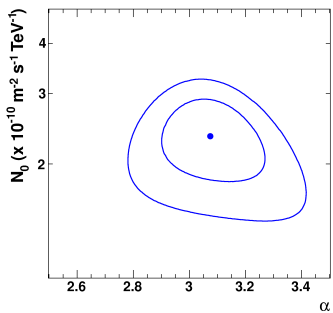}
\end{center}
\caption{MGRO J2031+41: 1 and 2-$\sigma$ C.L. contours for the power law fit. The point indicates the best fit
result (minimum $\chi^2$).
\label{f.2031contPL}}
\end{figure}

\begin{figure}
\begin{center}
\includegraphics[angle=0,width=7.5cm]{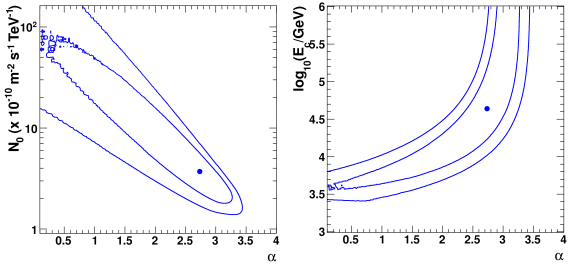}
\end{center}
\caption{MGRO J2031+41: 1 and 2-$\sigma$ C.L. contours for the power law with cutoff fit.
\textit{left panel} - $N_{10\mathrm{TeV}}$ vs $\alpha$ projection. \textit{right panel} - $E_c$ vs $\alpha$
projection. The points indicate the best fit result (minimum $\chi^2$).
The right panel shows that the present analysis is unable to constrain the upper limit of the energy cutoff
within 1$\sigma$.
\label{f.2031contCO}}
\end{figure}

Fig. \ref{f.2031contPL} and \ref{f.2031contCO} show the 1 and 2-$\sigma$ contours from the $\chi^2$
fit for the two tested hypotheses.

Best-fit spectral parameters are summarized in Table \ref{t.results}.

For the power law hypothesis, the best fit parameters are
$N_{10\mathrm{TeV}} = 2.4^{+0.6}_{-0.5} \cdot 10^{-10}$ $\mathrm{s^{-1}\: m^{-2}\: TeV^{-1}}$ and
$\alpha=3.08^{+0.19}_{-0.17}$.
The $\chi^{2}$ is 6.98 (7 d.o.f.), with a $\chi^{2}$ fit probability of 0.43.

\begin{figure}
\begin{center}
\includegraphics[angle=0,width=8cm]{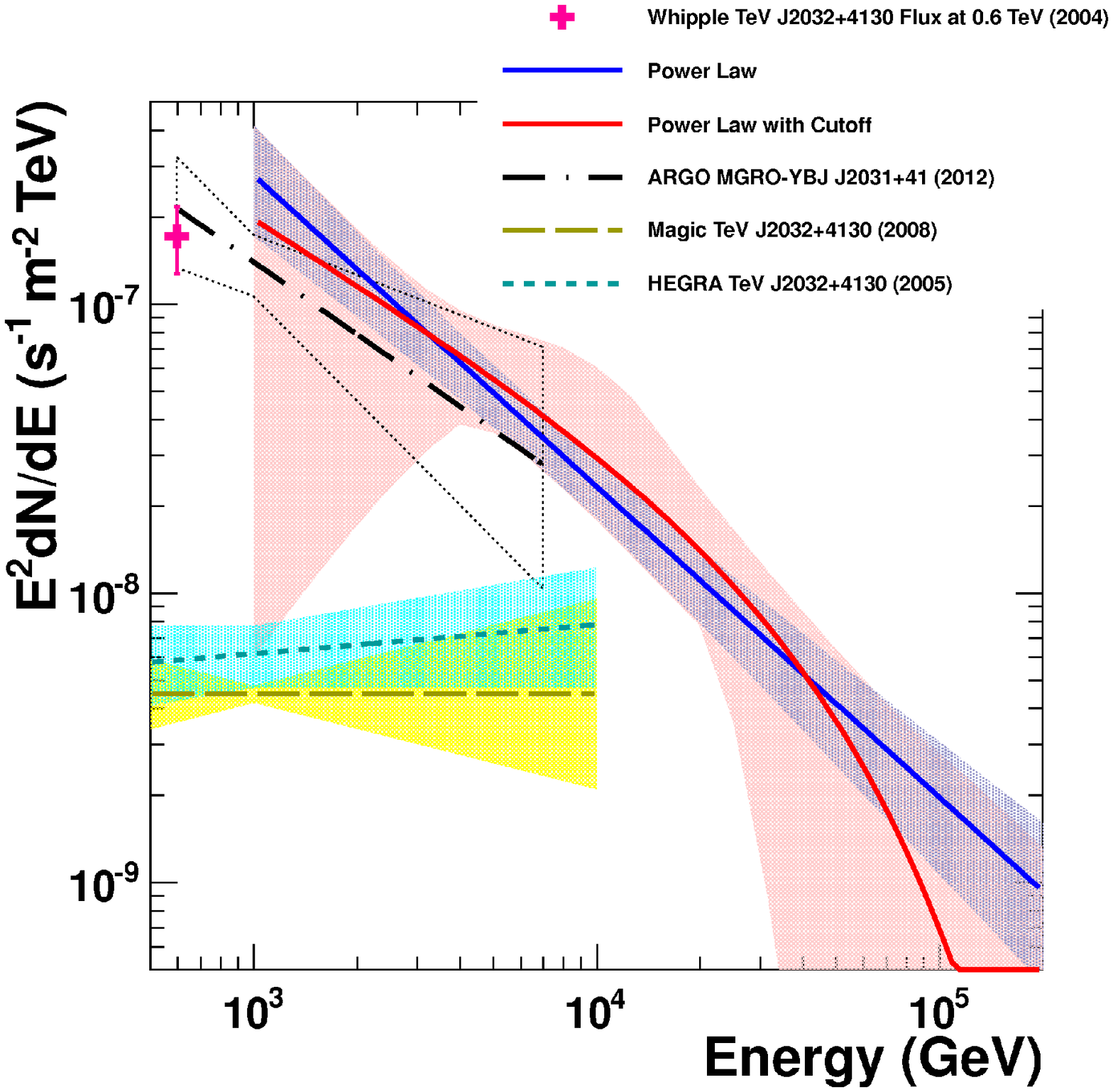}
\end{center}
\caption{MGRO J2031+41: energy spectra.
The power law model is shown in blue and the power law with cutoff model is shown in red.
These two hypotheses give the same $\chi^{2}$ fit probability.
The pink cross is the Whipple flux at 0.6 TeV \citep{whipple2004}. 
The fine-dashed cyan line and the dashed yellow line are the HEGRA and MAGIC
best fits \citep{hegra2005,magic2008}, respectively.
The dot-dash black line is the ARGO-YBJ best fit \citep{argo2012}.
The shadowed area represents the 1-$\sigma$ band.
\label{f.2031spectra}
}
\end{figure}

For the power law with cutoff hypothesis, the best fit parameters are 
$N_{10\mathrm{TeV}}=3.7^{+90.0}_{-1.9}\cdot10^{-10}$ $\mathrm{s^{-1}\: m^{-2}\: TeV^{-1}}$,
$\alpha=2.7^{+0.5}_{-2.6}$ and $E_c=44^{+\infty}_{-40}$ TeV.
In this case we are not able to constrain the upper limit of the 1-$\sigma$ uncertainty for the cutoff
energy (see right panel of Fig. \ref{f.2031contCO}).
The maximum $E_c$ simulated value is 1000 TeV, almost an order of magnitude above the
highest detected energy.
The $\chi^{2}$ is 5.80 (6 d.o.f.), with a $\chi^{2}$ fit probability of 0.45.
The best fit spectra for the two models are shown in Fig. \ref{f.2031spectra}.

The improvement in the fit obtained with the more complex model is not significant, according to the F-test,
and the simple power law is to be preferred over the power law with the exponential cutoff.

The flux of MGRO J2031+41 at 20 and 35 TeV, measured by previous Milagro analyses
\citep{milagro2007b,milagro2009}, is in agreement with the results presented here.
Results from ARGO-YBJ, MAGIC, HEGRA and Whipple \citep{argo2012,magic2008,hegra2005,whipple2004} are shown
in Fig. \ref{f.2031spectra}.
The HEGRA and MAGIC models are mutually consistent, but they disagree with our best fit
in terms of integral flux in the overlapping energy range.
Between 1 and 10 TeV the flux measured by MAGIC accounts for only $\sim$3\% of the flux measured by Milagro.
The spectrum as measured by the air Cherenkov telescopes (ACTs) is much harder ($\alpha\sim$2) than the Milagro power
law best fit ($\alpha\sim$3).
This discrepancy can be explained by the following two facts.
First, the angular resolution of HEGRA and MAGIC ($<0.1^{\circ}$) is much better than that of Milagro
($0.3^{\circ}$ to $0.7^{\circ}$).
The Whipple flux was measured at 0.6 TeV, with a PSF of $0.21^{\circ}$, and it agrees better with
the extrapolation of the Milagro result at lower energies.
The same is true for ARGO-YBJ, which, with an angular resolution between $0.47^{\circ}$ and $2.8^{\circ}$,
measured an emission consistent with the results presented here.
The measurement presented in this paper accounts for photons coming from a larger region around the nominal
source position compared to ACTs.
Second, the way the background is computed is also different.
Whipple used the ON/OFF method and observed the source with a significance of 3.3$\sigma$, while MAGIC primarily
was operated in wobble mode (5.6$\sigma$).
A possible explanation of the presented results is that TeV J2032+4130, whose extension is slightly larger than the
HEGRA and MAGIC angular resolution, is surrounded by an extended emission, $\sim$$4^{\circ}\times3^{\circ}$
according to the Milagro map (see Fig. \ref{f.maps} discussed in Sec. \ref{s.morphology}).
Therefore, Milagro, Whipple and ARGO-YBJ are not able to disentangle the extended emission from the central
source and observe a higher flux.
MAGIC and HEGRA, on the other hand, take into account the extended emission as a background.
As a result, the ACTs measure a fainter emission and a harder spectrum for TeV J2032+4130.

% MORPHOLOGY
\subsection{Morphology}\label{s.morphology}
\begin{figure*}
\begin{center}
\includegraphics[angle=0,width=10cm]{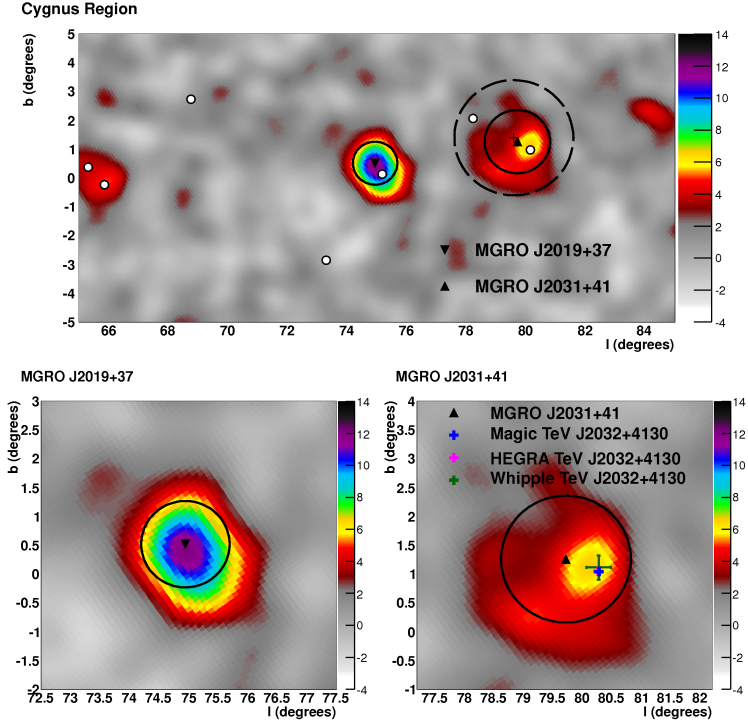}
\end{center}
\caption{PSF-smoothed significance maps of the Cygnus region.
\textit{top panel} - Black triangles mark the positions of the two Milagro sources with the 1-$\sigma$
radius (solid circles), white-filled circles mark Fermi BSL sources (the 1-$\sigma$ radius is consistent with the
marker size). The Fermi Cocoon position is marked by the cross and the dashed ellipse is its 1-$\sigma$ radius.
\textit{bottom left panel} - zoomed MGRO J2019+37 significance map. The circle represents the 1-$\sigma$ radius.
\textit{bottom right panel} - zoomed MGRO J2031+41 significance map. The Milagro position with
the 1-$\sigma$ radius is shown in black.
The crosses mark the Whipple (green), HEGRA (magenta) and MAGIC (blue) positions with their uncertainty.
\label{f.maps}}
\end{figure*}

The significance map of the Cygnus region for the last three years of Milagro data (2005-2008) is shown in
Fig. \ref{f.maps}.
An energy-dependent PSF smoothing is applied.
The top panel shows the entire region in Galactic coordinates.
Milagro positions for MGRO J2019+37 and MGRO J2031+41 are shown, as well as the high-significance
Fermi LAT sources ($>10\sigma$) and the extended Fermi Cocoon.

\begin{figure*}
\begin{center}
\includegraphics[angle=0,width=13.cm]{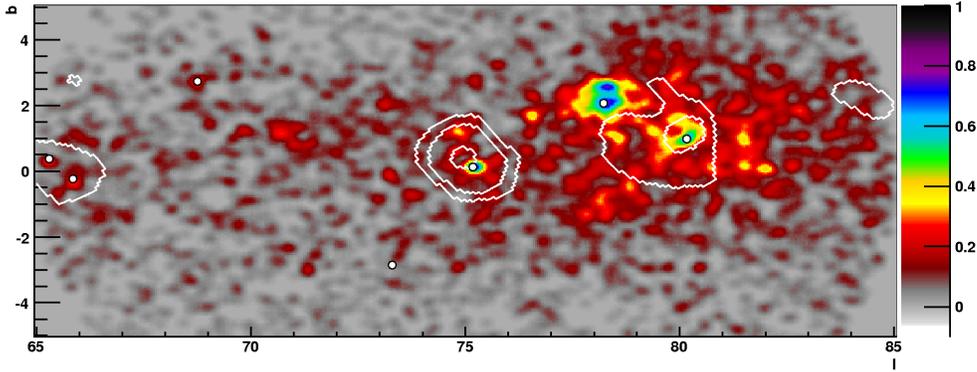}
\end{center}
\caption{Fermi LAT counts maps of the Cygnus region above 10 GeV.
We used Pass 7 photon data from 2008-08-04 15:43:37 to 2011-12-06 10:10:13 GMT \citep{latdata},
zenith angle $<107^{\circ}$, {\sc data\textunderscore qual}=1, {\sc lat\textunderscore config}=1.
A Gaussian smoothing is applied for display purposes.
Circles mark the position of the Fermi BSL sources.
White lines are the Milagro significance contours (3, 5 and 11$\sigma$). 
\label{f.fermi}}
\end{figure*}

The Fermi counts map for photon energies above 10 GeV overlaid with the Milagro significance contours in
the Cygnus region is shown in Fig. \ref{f.fermi}.
Milagro detects an excess consistent with the position of the two pulsars PSR J2021+3651 and
PSR J2032+4127, associated with MGRO J2019+37 and J2031+41 respectively, but no detection of
significant emission coincident with PSR J2021+4026.

MGRO J2019+37 is observed with a significance of 12.3$\sigma$.
Its extension is obtained with the 2-dimensional Gaussian fit discussed in Sec. \ref{s.results}.
The fit results in $\sigma=0.75^{\circ}$, slightly larger then the angular resolution of the detector in its
final configuration.
There is evidence from VERITAS that the extended source Milagro observes is the result of the superimposition
of close point-like sources \citep{veritas2011}.

MGRO J2031+41, observed with a significance of 7.3$\sigma$, is an extended source, with $\sigma=1.1^{\circ}$.
It shows a central core with a higher significance ($>6\sigma$) surrounded by a broader region
(approximately $3^{\circ}\times4^{\circ}$) with a lower significance ($>4\sigma$).
TeV J2032+4130 lies within the central core, and the Fermi Cocoon 1-$\sigma$ radius ($\sigma=2^{\circ}$)
completely includes MGRO J2031+41.
However, the Fermi analysis \citep{fermi2011b} shows that the Cocoon has an elongated shape, poorly
correlating with the Milagro contours, and the possible association between the Cocoon and the extended Milagro
excess remains unclear.

The spectral fit results do not change if the maximum significance position is chosen instead of the
2-D Gaussian fit position.

% CONCLUSIONS
\section{Conclusions}\label{s.conclusions}

We present the spectra of the two brightest Milagro sources in the Cygnus region using a
new analysis technique applied to the last 3 years of data collected by the Milagro experiment.

MGRO J2019+37 is observed with a significance over 12$\sigma$.
Its emission is well fitted by a power law with an exponential cutoff ($E_c=29^{+50}_{-16}$ TeV) and a hard
asymptotic spectral index ($\alpha=2.0^{+0.5}_{-1.0}$).
The simple power law hypothesis is disfavoured at a C.L. $>98$\%. 
The TeV excess measured by Milagro from MGRO J2019+37, spatially associated with the Fermi LAT pulsars J2018.0+3626
and J2021.0+3651, has been confirmed by VERITAS, and it is likely produced by nearby unresolved sources.
ARGO-YBJ does not detect a significant emission from MGRO J2019+37, but the 90\% C.L. upper limits do not
conflict with the Milagro best fitting model.

The emission from MGRO J2031+41 (7.3$\sigma$ significance), is well modeled by a power law with
$\alpha=3.08^{+0.19}_{-0.17}$.
Our result, in particular the integral flux in the overlapping energy range, is consistent with previous
measurements by ARGO-YBJ and Whipple, but it disagrees with HEGRA and MAGIC results for TeV J2032+4130, most
likely because of the different PSF of the instruments and different background subtraction methods.
MGRO J2031+41 appears to show a structured morphology, produced by the superimposition of a central point-like
source, coincident with TeV J2032+4130, and an extended emission, possibly produced either by unresolved sources
or interactions of cosmic rays with the local interstellar medium.
The correlation between the TeV extended emission and the overlapping Fermi Cocoon is unclear and needs
further studies.

HAWC, the next generation water Cherenkov observatory, will be able to produce a more accurate analysis of
the TeV emission from the Cygnus region, with its improved sensitivity (10-15 times better than Milagro) and
angular resolution.

% ACKNOWLEDGEMENTS
\acknowledgments
We gratefully acknowledge Scott Delay and Michael Schneider for their dedicated efforts in the construction and
maintenance of the Milagro experiment. This work has been supported by the National Science Foundation (under
grants PHY-0245234, -0302000, -0400424, -0504201, -0601080, -1002445, and ATM-0002744), the US Department of Energy
(Office of High-Energy Physics and Office of Nuclear Physics), Los Alamos National Laboratory, the University
of California, and the Institute of Geophysics and Planetary Physics.

\begin{flushleft}

\end{flushleft}

\clearpage

\begin{deluxetable}{lccc}
\tabletypesize{\scriptsize}
\tablecaption{Source Positions}
\tablewidth{0pt}
\tablehead{
  \colhead{survey} & \colhead{RA} & \colhead{Dec} & \colhead{width ($\sigma$)}\\
}
\startdata
MGRO J2019+37 &&&\\
2-D Gaussian with base          & 304.646$\pm$0.008 & 36.847$\pm$0.006 & 0.75 \\
Fermi position (0FGL J2020.8+3649)\tablenotemark{a} & 305.22            & 36.83            & ---  \\
Milagro 2007\tablenotemark{b}   & 304.98            & 36.66            & 1.1  \\
\hline\\
MGRO J2031+41 &&&\\
2-D Gaussian with base          & 307.41$\pm$0.02   & 41.190$\pm$0.015 & 1.1 \\
Fermi position (0FGL J2032.2+4122)\tablenotemark{a} & 308.06            & 41.38            & --- \\
Milagro 2007\tablenotemark{b}   & 307.98            & 41.51            & 3.0 \\
\enddata

\tablenotetext{a}{\citep{milagro2009}}
\tablenotetext{b}{\citep{milagro2007b}}
\label{t.positions}
\tablecomments{Positions of MGRO J2019+37 (top) and MGRO J2031+41 (bottom) for
the 2-D Gaussian with base fit, Fermi position of associated sources and previous Milagro
survey.}
\end{deluxetable}

\begin{deluxetable}{lcccccccccc}
\tabletypesize{\scriptsize}
\tablecaption{Fit Results}
\tablewidth{0pt}
\tablehead{
  \colhead{Source} & \colhead{$N_{10\mathrm{TeV}}$ $\mathrm{(10^{-10}\: s^{-1}\: m^{-2}\: TeV^{-1})}$} & \colhead{$\alpha$} & \colhead{$E_c$ $\mathrm{(TeV)}$} & $\chi^{2}/\mathrm{d.o.f.}$& \colhead{prob.}\\
}
\startdata
MGRO J2019+37 & $4.1\pm0.6$          & $2.78\pm0.10$          & $\infty$             & 16.12/7 & $0.024$ \\
MGRO J2019+37 & $7^{+5}_{-2}$      & $2.0^{+0.5}_{-1.0}$    & $29^{+50}_{-16}$     & 1.924/6 & $0.93$  \\
\hline\\
MGRO J2031+41 & $2.4^{+0.6}_{-0.5}$  & $3.08^{+0.19}_{-0.17}$ & $\infty$             & 6.980/7 & $0.43$  \\
MGRO J2031+41 & $3.7^{+90.0}_{-1.9}$ & $2.7^{+0.5}_{-2.6}$    & $44^{+\infty}_{-40}$ & 5.796/6 & $0.45$  \\
\enddata
\label{t.results}
\tablecomments{Best fit results for MGROJ 2019+37 and MGRO J2031+41 using both the power law and
the power law with cutoff hypotheses. The table lists statistical errors only, the systematic flux
error is $\sim$30\% and the systematic error on the spectral index is $\sim$0.1.
}
\end{deluxetable}

\clearpage

\end{document}